\documentclass[11pt]{article}

\usepackage[utf8]{inputenc}
\usepackage[T1]{fontenc}
\usepackage{tgtermes}
\usepackage[a4paper,margin=1in]{geometry}
\usepackage{amsmath,amssymb,amsthm}
\usepackage{booktabs}
\usepackage{multirow}
\usepackage{array}
\usepackage{tabularx}
\usepackage{microtype}
\usepackage{hyperref}
\usepackage{enumitem}
\usepackage{graphicx}
\usepackage{xcolor}
\usepackage{setspace}
\usepackage{caption}
\usepackage[numbers,sort&compress]{natbib}
\usepackage{url}
\usepackage{float}
\usepackage{subcaption}
\usepackage{cleveref}
\usepackage{siunitx}
\usepackage{pgfplots}
\usepackage{pgfplotstable}
\usepackage{algorithm}
\usepackage{algpseudocode}
\usepackage{titlesec}
\usepackage{titling}

\pgfplotsset{compat=1.18}
\usetikzlibrary{positioning}

\hypersetup{
  hidelinks,
  pdftitle={AI Inference as Relocatable Electricity Demand: A Latency-Constrained Energy-Geography Framework},
  pdfauthor={Xubin Luo; Cheng Yang}
}

\pretitle{\begin{center}\LARGE\bfseries}
\posttitle{\par\end{center}\vspace{0.5em}}
\preauthor{\begin{center}\normalsize}
\postauthor{\par\end{center}\vspace{-0.2em}}
\predate{\begin{center}\small}
\postdate{\par\end{center}\vspace{0.5em}\hrule\vspace{1.0em}}

\titleformat{\section}{\large\bfseries}{\thesection.}{0.5em}{}
\titleformat{\subsection}{\normalsize\bfseries}{\thesubsection}{0.5em}{}
\titlespacing*{\section}{0pt}{1.3em plus 0.2em minus 0.1em}{0.6em}
\titlespacing*{\subsection}{0pt}{1.0em plus 0.2em minus 0.1em}{0.4em}

\newtheorem{proposition}{Proposition}
\newtheorem{remark}{Remark}

\title{\textbf{AI Inference as Relocatable Electricity Demand:}\\
A Latency-Constrained Energy-Geography Framework}
\author{
Xubin Luo, Cheng Yang\textsuperscript{*}\\
\small School of Computer Science and Artificial Intelligence, Southwestern University of Finance and Economics, Chengdu, China\\
\footnotesize\textsuperscript{*}Corresponding author: Cheng Yang.
}
\date{Preprint, April 2026}

\begin{document}

\maketitle

\vspace{0.2em}

\begin{abstract}
AI inference is becoming a persistent and geographically distributed source of electricity demand. Unlike many traditional electrical loads, inference workloads can sometimes be executed away from the user-facing service location, provided that latency, state locality, capacity, and regulatory constraints remain acceptable. This paper studies when such digital relocation of computation can be interpreted as a latency-constrained relocation of electricity demand.

We develop an energy-geography framework for geo-distributed AI inference. The framework models a three-layer architecture of clients, service nodes, and compute nodes, and formulates inference placement as a constrained optimization problem over electricity prices, marginal carbon intensity, power usage effectiveness, compute capacity, network latency, and migration frictions. The key object is the energy--latency frontier: the marginal cost and carbon benefit unlocked by relaxing inference latency budgets.

The paper makes four contributions. First, it distinguishes physical electricity transmission from digital relocation of electricity-consuming computation, clarifying the limited but important sense in which inference can shift electricity demand across regions. Second, it formulates a geo-distributed inference placement model with legal/system feasibility masks and explicit migration frictions. Third, it introduces operational metrics---relocatable inference demand, energy return on latency, carbon return on latency, and a relocation break-even condition---to quantify when geographic relocation is meaningful. Fourth, it provides a transparent stylized simulation over representative global compute regions to illustrate how heterogeneous latency tolerance separates workloads into local, regional, and energy-oriented execution layers.

This study adopts a computational modeling approach to isolate the energy--latency mechanism. The simulation is designed to provide structural insight into feasible relocation patterns, rather than to estimate production-scale deployment magnitudes. The results show that latency relaxation expands feasible geography, while migration frictions, egress costs, state locality, legal constraints, and capacity limits can sharply reduce realized benefits.
\end{abstract}

\vspace{0.25em}

\section{Introduction}

AI energy discussions are shifting from training to inference. Training is episodic, but inference is recurrent and increasingly embedded in search, copilots, customer support, enterprise workflows, and agentic software. The relevant infrastructure question is therefore not only how much electricity AI consumes, but where that electricity is consumed.

This paper starts from a physical-digital asymmetry. Electricity moves through grids and is constrained by transmission topology, congestion, and capital-intensive infrastructure. Inference moves through communication networks and is constrained by latency, state locality, service-level objectives, and serving architecture. The two mechanisms are not equivalent, but they can be partially substitutable in operational terms.

We study the following research questions:
\begin{itemize}[leftmargin=1.5em]
    \item \textbf{RQ1:} Under what latency budgets can inference workloads be executed outside their local service region?
    \item \textbf{RQ2:} How much cost or carbon benefit is unlocked per unit of additional latency tolerance?
    \item \textbf{RQ3:} How do workload classes differ in relocatability once statefulness, migration friction, and capacity limits are included?
\end{itemize}

This paper does not claim that AI inference literally transmits electricity, nor that all workloads are globally mobile. The claim is narrower: when an inference request can be executed in multiple regions, placement determines where electricity-consuming computation is realized. For such workloads, latency, state, law, and system feasibility define a bounded relocation region.

Methodologically, we develop an energy-geography framework with three components: a formal optimization model, operational metrics, and a reproducible stylized simulation protocol. The model includes electricity price, marginal carbon signal, \(\mathrm{PUE}_i(t)\), capacity, latency, and migration frictions. The metrics include relocatable inference demand (\(\mathrm{RID}\)), energy return on latency (\(\mathrm{ERL}\)), carbon return on latency (\(\mathrm{CRL}\)), and break-even net benefit. The simulation is transparent and scenario-based rather than deployment-trace calibrated.

The paper makes four contributions. First, it provides a conceptual distinction between physical electricity movement and digital relocation of electricity-consuming computation. Second, it formulates a geo-distributed inference placement model with legal/system feasibility masks and explicit migration frictions. Third, it introduces measurable metrics and a break-even condition for when remote execution is meaningfully beneficial. Fourth, it offers a reproducible stylized simulation that demonstrates mechanism-consistent workload sorting into local, regional, and energy-oriented layers.

This preprint presents a conceptual and computational framework. Its quantitative results are used to clarify mechanism and boundary conditions rather than to infer production-scale magnitudes.

The remainder of the paper develops related work, the formal model, structural implications, simulation design, results, and limitations.

\section{Related Work}

This paper relates to several lines of work at the intersection of AI systems, geo-distributed computing, and energy-aware infrastructure. The most relevant strands include carbon-aware workload scheduling, geo-distributed service placement, energy-efficient LLM inference systems, and power-system perspectives on flexible digital loads. Our contribution differs from these literatures not primarily by introducing a new scheduling primitive, but by reframing AI inference as a geographically relocatable component of electricity demand and studying the resulting energy--latency tradeoff at the infrastructure level.

\subsection{Carbon-Aware AI Inference Routing}

A growing literature studies how geographically distributed computing resources can be used to reduce electricity cost or carbon emissions by shifting workloads across regions. In the AI context, this has become especially relevant as inference demand scales across cloud regions with heterogeneous electricity prices and carbon intensities.

One influential line of work studies carbon-aware routing of AI inference workloads. \citet{chien2023reducing} examine the carbon impact of generative AI inference and show that directing requests toward lower-carbon regions can reduce emissions substantially while preserving user experience. Their work highlights that inference, not just training, can dominate the long-run carbon burden of large-scale generative AI deployments.

Recent systems work extends this perspective. \citet{li2025ecoserve} propose EcoServe, a carbon-aware framework for AI inference systems that jointly considers operational and embodied emissions, using production traces from generative AI deployments to optimize inference provisioning and scheduling. The framework reports substantial carbon reductions under realistic service constraints. \citet{moore2025slit} formulate a multi-objective optimization problem for geo-distributed LLM scheduling that jointly considers time-to-first-token, carbon emissions, water use, and electricity cost, further expanding the sustainability objectives beyond carbon alone.

These studies establish that placement affects emissions and cost. Our departure is interpretive and measurable: rather than only optimizing where requests are served, we study when routing decisions operationally reassign the geography of electricity consumption.

\subsection{Geo-Distributed Load Balancing and Flexible Digital Loads}

Our work also connects to the long literature on geographical load balancing (GLB) in distributed systems. GLB studies how requests can be shifted across data centers to optimize latency, cost, reliability, or energy use. Foundational work showed that geo-distributed services can exploit spatial differences in electricity prices, renewable availability, and time-varying operating conditions to improve efficiency \citep{qureshi2009cutting,rao2010minimizing,liu2011greening,liu2012renewable}.

This literature provides the operational basis for our model: requests need not be served at the nearest or local compute site if the service architecture permits remote execution. However, most GLB work treats geographic diversity as an optimization opportunity within an already-established service model. The dominant question is how to route requests better, rather than how the interaction between service constraints and energy geography may shape the long-run form of inference infrastructure.

Our paper builds on the GLB intuition but asks a different question: when does geographically balancing inference become meaningful as relocation of electricity-demand realization, rather than only service optimization?

\subsection{Energy-Efficient LLM Serving}

Another closely related body of work focuses on reducing the energy footprint of LLM inference through system and architecture-level optimizations. These approaches typically optimize within a node or cluster rather than across a broad geographic network.

\citet{liu2025greenllm} study GPU energy management for LLM inference by dynamically controlling clock frequency across prefill and decode stages, thereby trading off energy savings against latency constraints. \citet{stojkovic2025dynamollm} dynamically reconfigure server instances and model-parallel execution to reduce operational carbon while maintaining latency targets. \citet{li2024sprout} reduce inference cost and carbon emissions by steering generation toward shorter responses, thereby lowering the number of autoregressive decoding steps. EcoServe \citep{li2025ecoserve} also belongs partly in this category because of its full-stack perspective on inference infrastructure and its treatment of embodied versus operational emissions.

These systems show the energy--latency tradeoff inside serving stacks. Our framework studies the same tradeoff across regions and asks when cross-region assignment remains meaningful after migration frictions and legal feasibility are internalized.

\subsection{Data Centers, Demand Response, and Flexible Digital Loads}

Outside the AI serving literature, prior work in energy systems and sustainable computing has explored data centers as flexible electrical loads. Data centers can participate in demand response, absorb surplus renewable generation, or shift workloads across sites in ways that support grid-level flexibility \citep{le2010capping,liu2012renewable}.

This literature is particularly important for our framing. It suggests that large computing infrastructures need not be treated as rigid demand sinks. Instead, they can function as controllable or partially movable loads, especially when some computation is delay-tolerant. Related work on virtual power lines, battery-backed load shifting, and flexible demand provides a useful conceptual backdrop: electricity systems often seek to reallocate where and when energy is consumed, whether via storage, transmission, or controllable demand.

Our contribution differs in the mechanism we emphasize. We do not study storage-based substitution or literal power-network virtualization. Instead, we focus on digital inference workloads whose execution site can be reassigned through communication networks, subject to latency and service constraints. Thus, the relevant comparison is not between physical and virtual transmission in the electrical sense, but between physical electricity movement and digital relocation of electricity-consuming computation.

\subsection{Gap and Positioning}

Prior work mostly asks how to route workloads more efficiently within distributed systems. This paper asks when such routing decisions become equivalent, in a limited operational sense, to relocating the geography of electricity consumption. The distinction is intentionally narrow: not physical-grid dispatch equivalence, but computation-location equivalence under latency, state, legal, and system constraints.

\subsection{How This Paper Differs}

Table~\ref{tab:related-positioning} conceptually summarizes the relationship between our paper and existing work.

\begin{table}[H]
\centering
\caption{Conceptual positioning of this paper relative to related work.}
\label{tab:related-positioning}
\small
\setlength{\tabcolsep}{6pt}
\renewcommand{\arraystretch}{1.15}
\begin{tabularx}{\textwidth}{
>{\raggedright\arraybackslash}p{2.9cm}
>{\raggedright\arraybackslash}X
>{\raggedright\arraybackslash}p{2.8cm}
>{\raggedright\arraybackslash}p{3.2cm}}
\toprule
\textbf{Line of work} & \textbf{Primary question} & \textbf{Typical objective} & \textbf{Spatial implication} \\
\midrule
Carbon-aware inference routing & Where should requests go to reduce emissions? & Carbon / cost / latency & Limited to request assignment \\
Geo-distributed load balancing & How should workloads be distributed across sites? & Latency / cost / utilization & Improves distributed service operation \\
Energy-efficient LLM serving & How can inference be made cheaper or greener inside a system? & Energy / latency / throughput & Mostly intra-node or intra-cluster \\
Flexible data-center loads & How can computing demand support the grid? & Grid flexibility / energy shifting & Data centers as controllable loads \\
\textbf{This paper} & How does latency-constrained inference relocation reshape where electricity is consumed? & Energy--latency frontier under SLOs & Hierarchical spatial geography of inference \\
\bottomrule
\end{tabularx}
\end{table}

In short, this paper is closest to carbon-aware and geo-distributed inference scheduling, but differs in its framing and level of analysis. We study AI inference not only as a schedulable workload, but as a geographically assignable form of electricity demand whose mobility is bounded by latency.

\section{Formal Model}

We consider a geo-distributed AI inference system in which inference requests can be routed across multiple compute locations with heterogeneous electricity prices, carbon intensities, and capacity limits. The objective is to characterize how service-level latency constraints interact with energy geography to determine the spatial allocation of inference workloads.

\subsection{System Architecture}

We model the system as a three-layer architecture:

\[
\text{Client} \rightarrow \text{Service Node} \rightarrow \text{Compute Node}
\]

A client \(u \in \mathcal{U}\) issues an inference request. The request is first received by a service node \(s \in \mathcal{S}\), which is responsible for request admission, orchestration, routing, and response streaming. The actual model execution occurs at a compute node \(i \in \mathcal{N}\), where GPUs or other accelerators consume electricity and perform inference.

This architecture reflects a common deployment pattern in modern AI services. User-facing service nodes are typically placed near demand centers, while inference clusters may be distributed across regions with different energy and infrastructure conditions.

\subsection{Task Model}

Let \(k \in \mathcal{K}\) denote an inference task. Each task is characterized by a tuple
\[
k = (u_k, s_k, \tau_k, E_k, D_k, m_k),
\]
where:
\begin{itemize}[leftmargin=1.5em]
    \item \(u_k\) is the originating client,
    \item \(s_k\) is the ingress service node,
    \item \(\tau_k\) is the latency tolerance or SLO budget,
    \item \(E_k\) is the estimated energy demand of the task,
    \item \(D_k\) is the compute demand,
    \item \(m_k\) is the number of service-to-compute interaction rounds.
\end{itemize}

The inclusion of \(m_k\) allows us to distinguish between simple single-shot inference and more complex request flows involving multiple backend interactions, such as tool use, retrieval-augmented generation, or iterative orchestration.

We allow heterogeneous task classes. In particular, interactive tasks typically have low \(\tau_k\), while batch or offline tasks may have large \(\tau_k\). This heterogeneity is central to the spatial allocation problem.

\subsection{Node Model}

Each compute node \(i \in \mathcal{N}\) is associated with time-varying energy and environmental attributes:
\[
P_i(t) = \text{electricity price at node } i \text{ at time } t,
\]
\[
\mathrm{MOER}_i(t) = \text{marginal operating emissions rate at node } i \text{ at time } t,
\]
\[
\mathrm{PUE}_i(t) = \text{power usage effectiveness at node } i \text{ at time } t,
\]
\[
\mathrm{Cap}_i(t) = \text{available compute capacity at node } i \text{ at time } t.
\]

Here \(P_i(t)\) is measured in monetary units per kWh, and \(\mathrm{MOER}_i(t)\) in \(gCO_2eq/kWh\). \(\mathrm{PUE}_i(t)\) maps accelerator-level energy to facility-level electricity demand. The capacity \(\mathrm{Cap}_i(t)\) may represent available GPU-seconds, token throughput, or another normalized compute resource.

Service nodes are assumed to be relatively close to users and primarily latency-oriented. Compute nodes, by contrast, may be located in regions optimized for energy cost, carbon profile, or infrastructure scale.

\subsection{End-to-End Latency Model}

For each task \(k\) assigned to compute node \(i\), the total latency is modeled as
\[
L_k(i) = L^{cs}_k + m_k \cdot L^{sc}_{s_k,i} + L^{q}_k(i) + L^{inf}_k(i),
\]
where:
\begin{itemize}[leftmargin=1.5em]
    \item \(L^{cs}_k\) is the client-to-service latency,
    \item \(L^{sc}_{s_k,i}\) is the one-round service-to-compute network latency between service node \(s_k\) and compute node \(i\),
    \item \(L^{q}_k(i)\) is the queueing or scheduling delay at node \(i\),
    \item \(L^{inf}_k(i)\) is the inference execution time at node \(i\).
\end{itemize}

We do not treat latency as a single network hop. Instead, the model isolates the service-to-compute component as the main geographically controllable variable. The client-to-service term is typically small and relatively stable under edge-oriented deployment, whereas \(L^{sc}_{s_k,i}\) grows with geographic distance and routing conditions.

A feasible assignment must satisfy the service-level constraint
\[
L_k(i) \le \tau_k.
\]

This constraint captures the core systems tradeoff: geographically distant compute nodes may offer lower electricity prices or lower carbon intensity, but they increase \(L^{sc}_{s_k,i}\) and may therefore violate the task's latency budget.

\subsection{Assignment Variables}

Let
\[
x_{k,i} \in \{0,1\}
\]
be a binary decision variable such that
\[
x_{k,i} =
\begin{cases}
1, & \text{if task } k \text{ is assigned to compute node } i,\\
0, & \text{otherwise.}
\end{cases}
\]

Each task must be assigned to exactly one compute node:
\[
\sum_{i \in \mathcal{N}} x_{k,i} = 1, \qquad \forall k \in \mathcal{K}.
\]

\subsection{Cost and Carbon Model}

We define \(E_k\) as accelerator-level task energy. If task \(k\) is assigned to node \(i\), facility-level electricity demand is
\[
\mathrm{FacilityEnergy}_{k,i} = E_k \cdot \mathrm{PUE}_i(t),
\]
its electricity cost is
\[
\mathrm{EnergyCost}_{k,i} = E_k \cdot \mathrm{PUE}_i(t)\cdot P_i(t),
\]
and its operational carbon impact is
\[
\mathrm{CarbonCost}_{k,i} = E_k \cdot \mathrm{PUE}_i(t)\cdot \mathrm{MOER}_i(t).
\]

These two quantities reflect the economic and environmental consequences of executing the same digital workload in different locations.

\subsection{Capacity Constraints}

The aggregate assigned demand at node \(i\) must not exceed its available capacity:
\[
\sum_{k \in \mathcal{K}} D_k \, x_{k,i} \le \mathrm{Cap}_i(t), \qquad \forall i \in \mathcal{N}.
\]

This constraint prevents the optimizer from unrealistically routing all requests to the cheapest or cleanest node.

\subsection{Latency Penalty}

In addition to the hard SLO constraint, it is useful to model soft latency pressure in the objective. Define a latency penalty as
\[
\mathrm{DelayPenalty}_{k,i} = g(L_k(i), \tau_k),
\]
where \(g(\cdot)\) is a nonnegative penalty function. A simple form is
\[
\mathrm{DelayPenalty}_{k,i} = \max\{0, L_k(i)-\tau_k\},
\]
if SLO violation is allowed but penalized, or alternatively
\[
\mathrm{DelayPenalty}_{k,i} = L^{sc}_{s_k,i},
\]
if the planner seeks to discourage long-distance routing even when hard feasibility holds.

In the baseline formulation below, we use hard SLO feasibility and a soft penalty on geographic latency to capture the intuition that extra distance consumes latency budget even before a violation occurs.

\subsection{Migration Frictions}

The baseline model treats each request as geographically assignable once latency and capacity permit. In practice, however, region switching incurs overhead. We decompose this friction as
\[
\mathrm{MigrationCost}_{k,i}
=
M^{state}_{k,i}+M^{cache}_{k,i}+M^{egress}_{k,i}+M^{replica}_{k,i}.
\]
This term captures state transfer, cache loss, networking, and artifact replication overhead:
\begin{itemize}[leftmargin=1.5em]
    \item \(M^{state}_{k,i}\): state transfer or replay overhead,
    \item \(M^{cache}_{k,i}\): extra prefill work caused by KV-cache loss,
    \item \(M^{egress}_{k,i}\): cross-region networking and egress charges,
    \item \(M^{replica}_{k,i}\): model or retrieval-data replication costs.
\end{itemize}

For stateless one-shot tasks, \(\mathrm{MigrationCost}_{k,i}\) may be small. For multi-turn or tool-augmented inference, it may overturn the apparent advantage of a remote low-price or low-carbon node. In autoregressive decoding, losing KV-cache affinity can require re-prefill or context recomputation, which raises both accelerator energy and end-to-end latency.

\subsection{Feasible Assignment Set}

We define legal and system feasibility masks \(a^{legal}_{k,i},a^{system}_{k,i}\in\{0,1\}\). The feasible compute set for task \(k\) is
\[
\mathcal{F}_k=\left\{i\in\mathcal{N}\ \middle|\ L_k(i)\le\tau_k,\ a^{legal}_{k,i}=1,\ a^{system}_{k,i}=1\right\}.
\]
Assignments outside \(\mathcal{F}_k\) are disallowed.

\subsection{Joint Optimization Problem}

The routing problem minimizes a weighted combination of electricity cost, carbon cost, latency penalty, and migration friction:
\begin{equation}
\begin{aligned}
\min_{\{x_{k,i}\}} \quad
& \sum_{k \in \mathcal{K}} \sum_{i \in \mathcal{N}} x_{k,i}
\Big(
\alpha E_k \mathrm{PUE}_i(t) P_i(t)
+ \beta E_k \mathrm{PUE}_i(t) \mathrm{MOER}_i(t) \\
& \qquad\qquad
+ \gamma \, \mathrm{DelayPenalty}_{k,i}
+ \eta \, \mathrm{MigrationCost}_{k,i}
\Big) \\
\text{s.t.} \quad
& \sum_{i \in \mathcal{N}} x_{k,i} = 1, \quad \forall k \in \mathcal{K}, \\
& \sum_{k \in \mathcal{K}} D_k x_{k,i} \le \mathrm{Cap}_i(t), \quad \forall i \in \mathcal{N}, \\
& x_{k,i}=0, \quad \forall k \in \mathcal{K},\ \forall i \notin \mathcal{F}_k, \\
& x_{k,i}\in \{0,1\}, \quad \forall k \in \mathcal{K},\ \forall i \in \mathcal{N}.
\end{aligned}
\label{eq:joint_optimization}
\end{equation}

Here \(\alpha,\beta,\gamma,\eta \ge 0\) are policy weights chosen by the operator. They encode the relative importance of economic cost, emissions, latency discipline, and migration friction. When \(\gamma\) or \(\eta\) is large, routing remains closer to users and pre-existing state. When \(\alpha\) or \(\beta\) dominate, the system becomes more willing to relocate tasks toward energy-favorable regions within \(\mathcal{F}_k\).

\subsection{Interpretive Implications}

This formulation makes explicit that AI inference is neither purely an energy optimization problem nor purely a latency minimization problem. It is a constrained spatial allocation problem. Energy-favorable locations create economic and environmental incentives for relocation, while latency budgets impose a geometric boundary on how far computation can move.

Under heterogeneous task classes, the model naturally implies differentiated routing outcomes. Tasks with small \(\tau_k\) are likely to remain in local or near-regional layers. Tasks with intermediate \(\tau_k\) can exploit regional diversity. Tasks with large \(\tau_k\) may migrate to energy-oriented compute locations far from users. This observation motivates the hierarchical spatial structure studied in the next section.

\subsection{Operational Metrics}

To make the framework testable and comparable across studies, we define several operational quantities.

\paragraph{Relocatable Inference Demand (\(\mathrm{RID}\)).}
\[
\mathrm{RID}=\frac{\sum_{k\in\mathcal{K}}E_k\cdot \mathbf{1}(i_k\neq i_k^{\text{local}})}{\sum_{k\in\mathcal{K}}E_k}.
\]

\paragraph{Energy Return on Latency (\(\mathrm{ERL}\)).}
For latency relaxation step \(\Delta\tau>0\),
\[
\mathrm{ERL}(\Delta\tau)=\frac{\mathrm{Cost}(\tau)-\mathrm{Cost}(\tau+\Delta\tau)}{\Delta\tau},
\qquad
\mathrm{CRL}(\Delta\tau)=\frac{\mathrm{Carbon}(\tau)-\mathrm{Carbon}(\tau+\Delta\tau)}{\Delta\tau}.
\]

\paragraph{Relocation break-even condition.}
Remote execution is meaningful only when it is feasible and net-beneficial. For candidate node \(i\),
\[
\mathrm{NB}_{k,i}=J_{k,local}-J_{k,i}>0,\quad i\in\mathcal{F}_k.
\]
Low electricity price or low carbon alone is insufficient if migration friction and latency penalties dominate.

\begin{table}[H]
\centering
\caption{Main notation used in the model.}
\label{tab:symbols}
\begin{tabular}{p{2.2cm} p{10.2cm}}
\toprule
\textbf{Symbol} & \textbf{Meaning} \\
\midrule
\(\mathcal{U}\) & set of clients \\
\(\mathcal{S}\) & set of service nodes \\
\(\mathcal{N}\) & set of compute nodes \\
\(\mathcal{K}\) & set of inference tasks \\
\(k\) & a task \\
\(i\) & a compute node \\
\(s_k\) & ingress service node for task \(k\) \\
\(\tau_k\) & latency tolerance of task \(k\) \\
\(E_k\) & estimated energy demand of task \(k\) \\
\(D_k\) & compute demand of task \(k\) \\
\(m_k\) & number of service-compute interaction rounds \\
\(P_i(t)\) & electricity price at node \(i\) and time \(t\) \\
\(\mathrm{MOER}_i(t)\) & marginal operating emissions rate at node \(i\) and time \(t\) \\
\(\mathrm{PUE}_i(t)\) & power usage effectiveness at node \(i\) and time \(t\) \\
\(\mathrm{Cap}_i(t)\) & available capacity at node \(i\) and time \(t\) \\
\(L^{cs}_k\) & client-to-service latency for task \(k\) \\
\(L^{sc}_{s_k,i}\) & service-to-compute latency from \(s_k\) to node \(i\) \\
\(L^{q}_k(i)\) & queueing delay of task \(k\) at node \(i\) \\
\(L^{inf}_k(i)\) & inference time of task \(k\) at node \(i\) \\
\(L_k(i)\) & total latency of task \(k\) if assigned to node \(i\) \\
\(x_{k,i}\) & binary assignment variable \\
\(\mathrm{MigrationCost}_{k,i}\) & Routing friction from state transfer, cache loss, egress, and replication. \\
\(a^{legal}_{k,i},a^{system}_{k,i}\) & legal and system feasibility masks for task-node assignment \\
\(\mathcal{F}_k\) & feasible compute set for task \(k\) \\
\(\mathrm{RID}\) & relocatable inference demand (energy-weighted) \\
\(\mathrm{ERL}, \mathrm{CRL}\) & energy and carbon return on latency relaxation \\
\(\mathrm{NB}_{k,i}\) & net benefit of remote node \(i\) relative to local execution \\
\(\alpha,\beta,\gamma,\eta\) & weights for energy, carbon, latency penalty, and migration friction \\
\bottomrule
\end{tabular}
\end{table}

\section{Conceptual Framework and Structural Implications}

The formulation in Section~3 implies that the spatial allocation of AI inference is governed by a structural asymmetry between physical electricity transmission and digital compute relocation. This section develops that asymmetry into an energy--latency tradeoff perspective and explains why it may induce a hierarchical geography of inference infrastructure.

\subsection{Physical Electricity Transmission vs.\ Digital Compute Relocation}

Electricity and AI inference are both linked to energy consumption, but they move through space under fundamentally different constraint regimes.

Physical electricity transmission is limited primarily by infrastructure cost, grid topology, congestion, and thermal capacity. Long-distance transmission often requires large capital investment, cross-jurisdiction coordination, and in some cases may be economically impractical, especially across oceans or weakly connected grid regions. Once electricity reaches a compute node, however, its geographic origin typically does not affect service latency or end-user experience.

By contrast, AI inference workloads are digitally routable. A request can in principle be executed at different compute nodes without physically transporting electricity. The marginal network transport cost of digital requests is often much lower than the infrastructure cost of expanding long-distance power transmission. However, digital relocation is not unconstrained. Its main limitation is latency: as compute is placed farther from the service layer or user-facing frontend, end-to-end service delay rises and may violate SLO requirements.

This leads to a functional asymmetry:
\begin{itemize}[leftmargin=1.5em]
    \item \textbf{electricity transmission} is mainly \textbf{cost-constrained},
    \item \textbf{compute relocation} is mainly \textbf{latency-constrained}.
\end{itemize}

The two mechanisms are not equivalent, but they are partially substitutable in the following sense: if a workload can be executed in a region with cheaper or cleaner electricity, then part of the electricity demand that would otherwise have been consumed near the user is instead consumed remotely. What is relocated is not electricity itself, but the location of electricity consumption. For this reason, we interpret geo-distributed AI inference as a mechanism for the \textbf{virtual relocation of electricity demand}, rather than literal power transmission.

\begin{figure}[H]
\centering
\resizebox{0.98\textwidth}{!}{%
\begin{tikzpicture}[
    x=1cm,y=1cm,
    every node/.style={font=\small},
    box/.style={
        draw,
        rounded corners=3pt,
        line width=0.5pt,
        align=center,
        minimum height=0.95cm,
        inner sep=5pt
    },
    compute/.style={
        draw,
        rounded corners=3pt,
        line width=0.5pt,
        align=center,
        minimum width=3.25cm,
        minimum height=0.82cm,
        inner sep=3pt,
        fill=green!8
    },
    arrow/.style={->, thick},
    dasharrow/.style={->, thick, dashed},
    lab/.style={
        font=\footnotesize,
        inner sep=1pt,
        align=center
    },
    tinylabel/.style={
        font=\scriptsize,
        inner sep=1pt,
        align=center
    }
]

\node[font=\bfseries\small, anchor=west] at (-0.2,3.65)
{Digital layer: inference requests};

\node[box, fill=blue!8, minimum width=2.45cm] (user) at (0,2.45)
{User\\regions};

\node[box, fill=blue!8, minimum width=2.65cm] (service) at (4.00,2.45)
{Service\\nodes};

\draw[arrow] (user.east) -- node[lab, above=4pt, pos=0.5] {requests} (service.west);

\node[font=\bfseries\small, anchor=west] at (7.05,3.65)
{Compute regions};

\node[compute, fill=green!5] (local) at (9.05,3.05)
{\textbf{Local}\\[-1pt]\scriptsize strict SLO};

\node[compute, fill=green!9] (regional) at (9.05,2.18)
{\textbf{Regional}\\[-1pt]\scriptsize moderate SLO};

\node[compute, fill=green!13] (energy) at (9.05,1.31)
{\textbf{Energy-oriented}\\[-1pt]\scriptsize relaxed SLO};

\draw[arrow] (service.east) -- node[tinylabel, above=4pt, sloped, pos=0.50]
{constrained routing} (local.west);

\draw[arrow] (service.east) -- (regional.west);

\draw[arrow] (service.east) -- (energy.west);

\node[font=\bfseries\small, anchor=west] at (-0.2,-0.85)
{Physical layer: electricity flow};

\node[box, fill=orange!10, minimum width=2.45cm] (power) at (0,-2.05)
{Power\\generation};

\node[box, fill=orange!8, minimum width=2.8cm] (grid) at (4.00,-2.05)
{Grid and\\transmission};

\node[box, fill=orange!12, minimum width=4.05cm] (consume) at (9.05,-2.05)
{Electricity consumed\\where inference runs};

\draw[arrow] (power.east) -- node[lab, above=4pt, pos=0.5] {electricity} (grid.west);

\draw[arrow] (grid.east) -- node[lab, above=6pt, pos=0.52]
{grid-constrained\\flow} (consume.west);

\draw[dasharrow] (energy.south) -- node[lab, right=5pt, pos=0.50]
{demand\\realized} (consume.north);

\end{tikzpicture}%
}
\caption{\mbox{Latency-constrained} relocation of inference demand. Digital routing assigns inference workloads to local, regional, or energy-oriented compute regions under latency, state, legal, capacity, and \mbox{migration-friction} constraints. Local, regional, and energy-oriented compute correspond respectively to \mbox{strict-SLO} interactive tasks, \mbox{moderate-SLO} online tasks, and \mbox{relaxed-SLO} batch or background workloads. The selected compute region determines where electricity demand is realized, while physical electricity remains constrained by the grid.}
\label{fig:conceptual-architecture}
\end{figure}

\subsection{The Energy--Latency Frontier}

For a given task \(k\), assigning execution to a more distant compute node may reduce electricity cost and carbon intensity while increasing service-to-compute latency. This creates a frontier between energy benefit and latency expenditure.

To illustrate, let the expected location-dependent cost of assigning task \(k\) to node \(i\) be
\[
J_{k,i}
=
\alpha E_k \mathrm{PUE}_i(t)P_i(t)
+
\beta E_k \mathrm{PUE}_i(t)\mathrm{MOER}_i(t)
+
\gamma \,\mathrm{DelayPenalty}_{k,i}
+
\eta \,\mathrm{MigrationCost}_{k,i}.
\]

This expression matches the formal objective in Section~3. For exposition, one may temporarily set migration friction aside, but the implemented joint policy includes it. If we ignore capacity constraints for the moment, the preferred execution node for task \(k\) is the feasible node with the smallest \(J_{k,i}\). As geographic distance increases, two opposing tendencies may emerge:
\begin{enumerate}[leftmargin=1.5em]
    \item \textbf{Energy advantage}: distant nodes may offer lower \(P_i(t)\) or lower \(\mathrm{MOER}_i(t)\), especially if they are located near stranded renewables, hydro-rich regions, or low-price wholesale markets.
    \item \textbf{Latency disadvantage}: distant nodes increase \(L^{sc}_{s_k,i}\), which consumes latency budget and raises the risk of infeasibility or degraded responsiveness.
\end{enumerate}

This tradeoff implies that the set of feasible low-cost assignments is bounded by task-specific latency tolerance. A task with strict \(\tau_k\) may only exploit local or nearby energy diversity. A task with relaxed \(\tau_k\) may access much broader geographic variation in price and carbon intensity.

We therefore define the \textbf{energy--latency frontier} as the set of optimal tradeoff points between:
\begin{itemize}[leftmargin=1.5em]
    \item achievable reduction in electricity cost or carbon cost, and
    \item latency budget consumed to obtain that reduction.
\end{itemize}

Operationally, one can study this frontier by varying latency tolerance or allowable geographic radius and measuring the resulting changes in total energy cost and emissions. The key quantity is not simply whether relocation is possible, but the \textbf{marginal gain from additional latency budget}:
\[
\frac{\partial \text{EnergySaving}}{\partial \tau},
\qquad
\frac{\partial \text{CarbonReduction}}{\partial \tau}.
\]

These derivatives capture how much additional cost or carbon benefit is unlocked when service constraints are relaxed by a small amount. In practice, this marginal gain is unlikely to be linear. In many systems, modest relaxations in latency tolerance may sharply expand the reachable set of compute nodes, after which further relaxation yields diminishing returns.

\subsection{Task Heterogeneity and Spatial Differentiation}

The energy--latency frontier is task-dependent. Different workloads consume latency budget in different ways and therefore exhibit different degrees of geographic flexibility.

Tasks with strong interactivity, such as real-time copilots, voice assistants, or highly responsive chat, typically have low \(\tau_k\) and often also small tolerance for service-to-compute round trips. These tasks are likely to remain close to service nodes or users.

Tasks with moderate responsiveness requirements, such as standard chat, enterprise question answering, or non-streaming assistant workflows, may tolerate larger \(L^{sc}_{s_k,i}\) and can therefore exploit regional diversity in electricity price and carbon intensity.

Weakly real-time or offline tasks, such as batch document analysis, background agent execution, asynchronous summarization, or synthetic data generation, may have very large \(\tau_k\). For these workloads, latency imposes only a weak constraint, and the optimization becomes increasingly dominated by price, carbon, and capacity.

Thus, task heterogeneity transforms the spatial allocation problem from a single placement decision into a differentiated sorting mechanism. Tasks are naturally stratified according to latency sensitivity, and each stratum has access to a different subset of the energy landscape.

\begin{proposition}[Monotonic feasible-set expansion]
For task \(k\), define \(\mathcal{F}_k(\tau_k)=\{i\in\mathcal{N}\mid L_k(i)\le\tau_k,\ a^{legal}_{k,i}=1,\ a^{system}_{k,i}=1\}\). If non-latency feasibility terms are fixed, then \(\tau_k' \ge \tau_k\) implies \(\mathcal{F}_k(\tau_k)\subseteq \mathcal{F}_k(\tau_k')\). Therefore, under unchanged policy weights and without additional capacity interactions, the minimum attainable objective value is weakly non-increasing as \(\tau_k\) increases.
\end{proposition}

\begin{remark}
Feasible-set expansion does not guarantee actual relocation. Sorting toward energy-favorable nodes occurs only when remote net benefit remains positive after migration frictions and capacity effects are internalized.
\end{remark}

\subsection{Hierarchical Spatial Structure}

Because task classes differ in latency tolerance while compute nodes differ in energy characteristics, the model suggests that AI inference infrastructure may evolve toward a \textbf{hierarchical spatial structure} rather than a flat, uniform deployment.

We hypothesize three broad layers:

\paragraph{Local Layer.}
The local layer contains compute capacity placed close to users or service nodes. Its primary purpose is to serve highly latency-sensitive workloads. These nodes may face high electricity prices and stricter local constraints, but they minimize service latency.

\paragraph{Regional Layer.}
The regional layer consists of major data-center hubs that balance moderate latency with economies of scale. These locations are typically farther from users than local edge-adjacent resources but still sufficiently connected to serve interactive or semi-interactive workloads. Many standard online AI services are likely to concentrate here.

\paragraph{Energy-Oriented Layer.}
The energy-oriented layer consists of compute locations selected primarily for favorable electricity price, carbon profile, or large-scale infrastructure conditions. These may include regions with abundant renewables, low wholesale electricity prices, or lower cooling and land costs. Such nodes are most suitable for weakly real-time and offline workloads.

This layered structure is not imposed \emph{a priori} in the optimization model. Rather, it emerges as a plausible macroscopic consequence of heterogeneous latency constraints and heterogeneous energy geography. In other words, the model suggests that AI infrastructure may be shaped jointly by user proximity, regional aggregation, and remote energy advantage.

\subsection{A Relocation View of Electricity Demand}

Under this interpretation, geo-distributed inference does not merely move computation; it changes \textbf{where electricity is consumed}. A request that would otherwise have triggered GPU execution in a high-price urban region may instead consume electricity in a distant region with lower price or lower carbon intensity, provided the resulting latency remains acceptable.

This motivates a relocation view of AI inference:
\begin{itemize}[leftmargin=1.5em]
    \item traditional industrial electricity demand is geographically anchored,
    \item AI inference makes part of electricity demand geographically assignable,
    \item latency and SLOs define the feasible relocation boundary.
\end{itemize}

The consequence is not that digital networks replace physical power grids. Rather, digital networks create an additional mechanism through which electricity demand can be reallocated across space. This mechanism may complement transmission, storage, and demand response by shifting selected forms of electricity consumption toward more favorable energy regions.

\subsection{Scope of the Claim}

The scope of the claim is limited to workloads whose execution location remains technically and legally assignable. State locality, data residency, egress charges, queueing uncertainty, and hardware heterogeneity shrink the feasible relocation set. Within that constrained set, the model characterizes how latency tolerance and energy heterogeneity jointly shape geographic placement.

This perspective provides the conceptual foundation for the reference evaluation specification that follows.

\section{Stylized Simulation Design}

This section specifies an evaluation protocol \emph{consistent with} the model in Section~3. It fixes notation for data and procedures, lists baselines and metrics, and makes explicit what a quantitative study should measure.

The specification imagines a simulation framework that combines geographically distributed compute nodes, time-varying energy conditions, and heterogeneous inference workloads. The goal is not to reproduce every detail of a production AI serving stack, but to describe how latency tolerance would affect the feasible relocation of inference demand across space once the model is instantiated with data.

The present manuscript implements a reduced-form version of this specification. The implementation focuses on the core mechanisms required to examine the energy--latency frontier, workload-class sorting, and migration-friction effects. Several production-level extensions---such as measured carbon traces, dynamic queueing, full legal routing masks, and heterogeneous GPU fleet modeling---are left for future empirical work. We refer to this implementation as a \emph{partially anchored stylized simulation}: inter-region latency and network-cost inputs are anchored to public cloud references, while carbon profiles, workload shares, and capacity assumptions remain stylized.

\subsection{Evaluation Objectives}

A stylized simulation following this specification is organized around four questions.

First, how much inference demand can be geographically relocated as task-level latency tolerance increases?

Second, how much electricity cost and carbon cost can be reduced by such relocation under realistic heterogeneity in price and carbon intensity?

Third, how do different task classes sort across local, regional, and energy-oriented compute layers?

Fourth, under what conditions does the system exhibit diminishing returns, that is, a point beyond which additional latency budget yields limited extra energy or carbon benefit?

These questions articulate what the simulation must examine to assess the paper's core claim that AI inference behaves as a latency-constrained mechanism for relocating electricity demand.

\subsection{Geographic Node Set}

We model a set of candidate compute nodes distributed across major global infrastructure regions. Each node represents a data-center cluster or metropolitan compute hub rather than an individual facility. The initial evaluation uses a moderate number of nodes so that the sources of variation remain interpretable.

A representative node set includes:
\begin{itemize}[leftmargin=1.5em]
    \item Virginia
    \item Oregon
    \item Frankfurt
    \item London
    \item Singapore
    \item Tokyo
    \item Dubai
    \item Sydney
    \item Beijing
    \item S\~ao Paulo
\end{itemize}

These nodes are chosen to capture diversity along three dimensions:
\begin{enumerate}[leftmargin=1.5em]
    \item \textbf{user proximity}, since some are major service-demand centers;
    \item \textbf{network geography}, since they span North America, Europe, Asia-Pacific, the Middle East, and South America;
    \item \textbf{energy geography}, since they face different electricity prices and carbon profiles.
\end{enumerate}

In later sensitivity tests, the node set can be expanded to additional regions, but the initial design prioritizes interpretability over raw scale.

\begin{table}[H]
\centering
\caption{Input status in the reduced-form simulation.}
\label{tab:node-parameter-template}
\small
\begin{tabularx}{\textwidth}{p{3.1cm}X}
\toprule
\textbf{Input category} & \textbf{Status in current simulation} \\
\midrule
Inter-region latency & Anchored to Azure region-to-region RTT where available; distance-based fallback for missing pairs. \\
Network transfer cost & Anchored to public Azure inter-region egress pricing references. \\
Electricity price & Partially anchored through market-linked or regional proxies. \\
Carbon intensity & Stylized hourly variation using regional carbon-profile proxies. \\
\(\mathrm{PUE}_i(t)\) and capacity & Assumed constant or share-based parameters for structural simulation. \\
Workload mix & Stylized task-class assumptions rather than production traces. \\
\bottomrule
\end{tabularx}
\end{table}

\subsection{Electricity Price and Carbon Data}

For each node \(i\), we associate two time-varying exogenous signals:
\[
P_i(t) = \text{electricity price},
\qquad
\mathrm{MOER}_i(t) = \text{marginal operating emissions rate}.
\]

Electricity price data are constructed from publicly available wholesale electricity market sources when possible, and otherwise from region-level proxies consistent with local market conditions. Carbon intensity data are obtained from grid-level carbon estimates or corresponding public API sources.

For a stronger quantitative instantiation, the preferred empirical path is to combine three observable inputs: (i)~measured inter-region latency or RTT matrices from cloud-region probing; (ii)~hourly regional electricity-price series from wholesale markets or regulated proxies; and (iii)~hourly carbon-intensity or marginal-emissions signals from grid-carbon datasets such as balancing-authority or ISO-level estimates. The current simulation only partially meets that standard: some electricity-price anchors are drawn from observable series or market-linked proxies, while several node-level trajectories remain stylized. This distinction matters because it clarifies what the present numbers can and cannot claim.

\paragraph{Input status in the current implementation.}
The present implementation uses a mixed input stack rather than a fully empirical one. Inter-region latency is anchored to an Azure region-to-region RTT reference, and inter-region transfer cost is anchored to public Azure bandwidth pricing. Electricity prices are partially anchored through regional proxies and scaling, while hourly carbon variation remains stylized unless an external carbon-intensity trace is supplied. Workload-class parameters, demand shares, and capacity shares are illustrative structural assumptions rather than measured deployment traces.

The model operates at an hourly granularity. This choice is deliberate. It is fine-grained enough to capture meaningful variation in electricity price and emissions signals while still supporting tractable large-scale simulations. For each experiment horizon, the system observes synchronized time series \(\{P_i(t), \mathrm{MOER}_i(t)\}\) across all nodes and solves routing decisions for the workload arriving at each interval.

We do not assume that operators have perfect long-horizon foresight. In the baseline setting, routing uses contemporaneous or short-horizon observable price and carbon signals. A robustness extension can later test imperfect prediction or delayed information.

\subsection{Latency Estimation}

Baseline simulation uses measured or published inter-region RTT values where available (repository RTT matrix). For each pair of service node \(s\) and compute node \(i\), we convert RTT to an effective service-to-compute latency term under the scenario's WAN inflation assumption.

When a pair is missing in the RTT matrix, we use a distance fallback:
\[
L^{sc}_{s,i}
\approx
\frac{d(s,i)}{v_f} + \delta,
\]
where:
\begin{itemize}[leftmargin=1.5em]
    \item \(d(s,i)\) is the great-circle distance between \(s\) and \(i\),
    \item \(v_f\) is the effective propagation speed in fiber,
    \item \(\delta\) is a fixed overhead term capturing routing, switching, and protocol overhead.
\end{itemize}

In the baseline implementation, this becomes
\[
L^{sc}_{s,i} \approx \frac{d(s,i)}{200 \text{ km/ms}} + 20 \text{ ms}.
\]

The fallback is not intended to recover exact Internet RTTs. It is a consistent approximation for missing links only. Sensitivity tests vary overhead and multiplicative inflation factors to represent suboptimal routing and congestion. This keeps the baseline anchored to measured RTT while preserving complete connectivity for stylized analysis.

\subsection{Task Classes}

A central feature of the evaluation is workload heterogeneity. Rather than sampling all tasks from a single latency distribution, we divide inference requests into four classes.

\paragraph{Class A: Interactive Tasks.}
Examples include real-time chat, copilots, and low-latency assistant interactions. These tasks have strict latency budgets and limited tolerance for remote execution.

Typical attributes:
\begin{itemize}[leftmargin=1.5em]
    \item low \(\tau_k\),
    \item small to moderate \(E_k\),
    \item low \(m_k\),
    \item high penalty for latency expansion.
\end{itemize}

\paragraph{Class B: Standard Online Tasks.}
Examples include standard chatbot requests, enterprise Q\&A, and general-purpose online inference. These tasks remain latency-sensitive, but are more tolerant than Class A.

Typical attributes:
\begin{itemize}[leftmargin=1.5em]
    \item moderate \(\tau_k\),
    \item moderate \(E_k\),
    \item moderate \(m_k\).
\end{itemize}

\paragraph{Class C: Weakly Real-Time Background Tasks.}
Examples include background agent workflows, asynchronous processing, multi-step enterprise automation, or deferred document generation.

Typical attributes:
\begin{itemize}[leftmargin=1.5em]
    \item large \(\tau_k\),
    \item moderate to high \(E_k\),
    \item higher \(m_k\) possible,
    \item relatively large geographic flexibility.
\end{itemize}

\paragraph{Class D: Offline Batch Tasks.}
Examples include batch summarization, synthetic data generation, document analytics, model distillation support tasks, and large-scale asynchronous pipelines.

Typical attributes:
\begin{itemize}[leftmargin=1.5em]
    \item very large \(\tau_k\),
    \item moderate to high \(E_k\),
    \item minimal responsiveness requirements.
\end{itemize}

\begin{table}[H]
\centering
\caption{Workload classes and representative parameter ranges used in stylized simulation.}
\label{tab:workload-classes}
\small
\begin{tabular}{p{1.7cm} p{3.4cm} p{2.9cm} p{2.4cm} p{2.5cm}}
\toprule
\textbf{Class} & \textbf{Example} & \textbf{Latency budget} & \textbf{Rounds \(m_k\)} & \textbf{Statefulness} \\
\midrule
A & interactive chat/copilot & 100--300\,ms & 1--2 & high \\
B & standard online QA & 500--1500\,ms & 1--3 & medium \\
C & background agents / async workflows & 5--60\,s & 3--10 & medium \\
D & offline batch inference & minutes to hours & variable & low \\
\bottomrule
\end{tabular}
\end{table}

Each task class is parameterized by distributions over:
\begin{itemize}[leftmargin=1.5em]
    \item latency tolerance \(\tau_k\),
    \item compute demand \(D_k\),
    \item energy demand \(E_k\),
    \item interaction depth \(m_k\),
    \item inference duration \(L^{inf}_k\).
\end{itemize}

This design allows the simulation to report not only average system behavior, but also how different workloads sort into different compute layers.

\subsection{Service Layer and Demand Distribution}

We model service nodes as ingress points located near major user demand centers. Each incoming task is first attached to a service node according to user geography. The service node then chooses among feasible compute nodes.

In the baseline setting, user demand is distributed across service nodes according to configurable regional weights. A simple initial setup may place larger demand on nodes such as Virginia, London, Singapore, Tokyo, and Beijing, while smaller demand originates from other regions. The precise distribution can be varied in robustness tests.

This separation between service nodes and compute nodes is important. It prevents the model from collapsing into a simple user-to-data-center assignment and better reflects the architecture of real AI services, in which frontend ingress and backend inference may be geographically distinct.

\subsection{Capacity Model}

Each compute node \(i\) is assigned a finite capacity \(\mathrm{Cap}_i(t)\). Capacity may be modeled in one of two ways:
\begin{enumerate}[leftmargin=1.5em]
    \item \textbf{static capacity}, where each node has a fixed upper bound over the evaluation period;
    \item \textbf{time-varying capacity}, where available capacity changes due to background utilization, maintenance, or other demand.
\end{enumerate}

The baseline evaluation uses static capacity for clarity. This allows us to isolate the interaction between latency and energy geography before introducing cluster volatility.

To avoid unrealistic concentration of all requests in the cheapest or cleanest region, node capacities are calibrated so that no single node can absorb the entire global workload.

\subsection{Baselines}

To assess the value of the proposed joint routing logic, we compare against several baselines. The list below is the baseline set for the stylized simulation and for future empirical extensions.

\paragraph{Local-Only.}
Each task is executed at the nearest local or service-adjacent compute node. This baseline minimizes latency but ignores energy geography.

\paragraph{Nearest-Region.}
Each task is executed at the geographically nearest feasible regional compute node. This captures conventional region-based deployment.

\paragraph{Price-Only.}
Each task is routed to the lowest-price feasible node, ignoring carbon except insofar as it enters feasibility through capacity or latency.

\paragraph{Carbon-Only.}
Each task is routed to the lowest-carbon feasible node, ignoring electricity price.

\paragraph{Joint Energy--Latency Policy.}
This is the proposed policy, which optimizes the weighted objective over electricity price, carbon intensity, and latency-related penalties subject to SLO and capacity constraints.

These baselines allow us to distinguish the effect of geographic energy awareness from pure latency minimization or single-objective routing.

\begin{table}[H]
\centering
\caption{Policy-weight normalization in the stylized baseline.}
\label{tab:policy-weights}
\small
\begin{tabular}{llll}
\toprule
\textbf{Component} & \textbf{Symbol} & \textbf{Baseline weight} & \textbf{Normalization} \\
\midrule
Electricity cost & \(\alpha\) & 1 & local-only task cost \\
Carbon term & \(\beta\) & 1 & local-only task carbon \\
Latency penalty & \(\gamma\) & 1 & class-specific latency scale \\
Migration friction & \(\eta\) & 1 & scenario-specific friction scale \\
\bottomrule
\end{tabular}
\end{table}

The normalized implementation evaluates the joint policy using
\[
\widetilde{J}_{k,i}
=
\alpha\frac{\mathrm{EnergyCost}_{k,i}}{\mathrm{EnergyCost}_{k,local}}
+\beta\frac{\mathrm{CarbonCost}_{k,i}}{\mathrm{CarbonCost}_{k,local}}
+\gamma\frac{\mathrm{DelayPenalty}_{k,i}}{\mathrm{DelayScale}_k}
+\eta\frac{\mathrm{MigrationCost}_{k,i}}{\mathrm{FrictionScale}_k}.
\]
The baseline uses equal weights after normalization. Weight sensitivity is treated as a robustness dimension rather than as a tuned source of the reported results.

\begin{table}[H]
\centering
\caption{Ablation and sensitivity scenario menu.}
\label{tab:ablation-menu}
\small
\begin{tabular}{lll}
\toprule
\textbf{Scenario group} & \textbf{Switch or variation} & \textbf{Status} \\
\midrule
Migration friction & off / on / high & reported in Section~6 \\
Egress cost & off / on & implemented as friction component \\
Carbon signal & average-intensity proxy / \(\mathrm{MOER}\) proxy & stylized proxy \\
Capacity regime & loose / baseline / tight & sensitivity menu \\
Workload mix & interactive-heavy / balanced / batch-heavy & sensitivity menu \\
Legal feasibility mask & off / on & future extension \\
\bottomrule
\end{tabular}
\end{table}

\subsection{Metrics}

We evaluate all policies using the following metrics. The current implementation reports the subset most directly tied to the energy--latency frontier, workload sorting, and migration-friction mechanisms.

\paragraph{Relocatable Inference Demand (\(\mathrm{RID}\)).}
The energy-weighted share of inference demand assigned to a node different from the local or nearest-region default:
\[
\mathrm{RID} = \frac{\sum_k E_k\, \mathbf{1}(i_k \neq i_k^{\text{local}})}{\sum_k E_k}.
\]

This measures how much demand becomes geographically mobile under each policy.

\paragraph{Electricity Cost Reduction.}
Relative reduction in total electricity expenditure compared to a baseline:
\[
\Delta \mathrm{Cost} =
\frac{\mathrm{Cost}^{\text{baseline}} - \mathrm{Cost}^{\text{policy}}}{\mathrm{Cost}^{\text{baseline}}}.
\]

\paragraph{Carbon Reduction.}
Relative reduction in total operational carbon:
\[
\Delta \mathrm{Carbon} =
\frac{\mathrm{Carbon}^{\text{baseline}} - \mathrm{Carbon}^{\text{policy}}}{\mathrm{Carbon}^{\text{baseline}}}.
\]

\paragraph{SLA Violation Rate.}
The share of tasks whose realized latency exceeds their tolerance:
\[
\mathrm{V} =
\frac{\sum_k \mathbf{1}(L_k > \tau_k)}{|\mathcal{K}|}.
\]

This is critical because a policy that reduces electricity cost at the expense of systematic SLO failure is not acceptable.

\paragraph{Tier Allocation Share.}
The fraction of workload assigned to local, regional, and energy-oriented layers, respectively. This metric is central to testing the hierarchical-structure hypothesis.

\paragraph{Energy Return on Latency (\(\mathrm{ERL}\)).}
For controlled sweeps over latency tolerance, we compute the incremental savings unlocked by additional latency budget:
\[
\frac{\Delta \mathrm{Cost}}{\Delta \tau},
\qquad
\frac{\Delta \mathrm{Carbon}}{\Delta \tau}.
\]

This reveals whether the system exhibits high early returns, smooth tradeoffs, or diminishing returns.

\subsection{Experimental Procedure}

The evaluation proceeds in three stages.

For each hour \(t\), the simulation constructs node-level price, carbon, and capacity conditions, then evaluates each workload class \(k\) at each source region against the feasible compute set implied by its latency budget. Candidate assignments that violate the end-to-end latency constraint are removed. The remaining feasible nodes are ranked under the Local-only, Price-only, or Joint policy, and workload is allocated subject to node capacity. Hourly allocations are then aggregated over the simulation horizon to compute total cost, total carbon, relocatable-demand share, and class-specific tier shares.

\paragraph{Stage 1: Latency Sweep.}
We vary task-level latency tolerance or a system-wide latency multiplier and measure how relocatable demand, cost savings, and carbon savings change. This produces the energy--latency frontier.

\paragraph{Stage 2: Task-Class Analysis.}
We hold the external node environment fixed and compare how different task classes are allocated under each routing policy. This reveals whether heterogeneous workloads naturally separate into different compute layers.

\paragraph{Stage 3: Sensitivity Analysis.}
We vary:
\begin{itemize}[leftmargin=1.5em]
    \item node capacity,
    \item latency inflation factors,
    \item electricity price volatility,
    \item carbon volatility,
    \item workload composition,
    \item policy weights \(\alpha,\beta,\gamma,\eta\).
\end{itemize}

These tests determine whether the observed spatial structure is robust or highly sensitive to modeling assumptions.

\subsection{Simulation Outputs}

The implementation produces three main outputs: an energy--latency frontier, workload-class tier allocation, and top relocation corridors. These outputs are reported in Section~6.

\subsection{Scope and Limitations of the Specification}

Any instantiation of this protocol will intentionally abstract away several practical factors, including state locality, detailed cloud pricing structures, data sovereignty rules, and exact WAN routing behavior. These omissions delimit what can be inferred from a given simulation run. The specification is meant to foreground the infrastructure-level energy--latency mechanism, not to reproduce every operational detail of a production serving platform.

A richer production model could later incorporate:
\begin{itemize}[leftmargin=1.5em]
    \item egress charges,
    \item retrieval locality,
    \item KV-cache affinity,
    \item token-level streaming effects,
    \item heterogeneous GPU fleets,
    \item water consumption,
    \item long-run data-center siting decisions.
\end{itemize}

For the conceptual aims of this manuscript, the simplified specification is sufficient to pose whether heterogeneous AI workloads, heterogeneous energy geography, and latency constraints can in principle induce differentiated spatial placement under transparent assumptions.

\section{Stylized Simulation Results and Structural Interpretation}

This section reports results from the reduced-form simulation. Rather than forecasting global deployment magnitudes, it tests whether the formal mechanism generates three expected patterns: an energy--latency frontier, workload-class sorting, and sensitivity to migration friction.

\subsection{Energy--Latency Frontier}

\Cref{fig:frontier} plots the medium-load scenario under a 1.4$\times$ WAN inflation factor, sweeping the system-wide latency multiplier and recording the resulting cost reduction, carbon reduction, and relocatable-demand fraction relative to the Local-only baseline. The simulation combines Azure region-to-region RTT measurements and Azure inter-region egress references with stylized hourly carbon variation.

\begin{figure}[H]
\centering
\begin{tikzpicture}
\begin{axis}[
    width=0.92\textwidth,
    height=7cm,
    xlabel={System-wide latency multiplier},
    ylabel={Percent relative to Local-only baseline},
    xmin=0.45, xmax=2.55,
    ymin=0, ymax=90,
    xtick={0.5,0.75,1.0,1.5,2.5},
    grid=both,
    grid style={line width=.1pt, draw=gray!20},
    major grid style={line width=.2pt,draw=gray!35},
    legend style={at={(0.03,0.97)}, anchor=north west, draw=none, fill=white, font=\small},
    tick label style={font=\small},
    label style={font=\small}
]
\addplot[color=blue, mark=o, thick] coordinates {
    (0.5,7.60) (0.75,9.13) (1.0,9.22) (1.5,9.92) (2.5,7.97)
};
\addlegendentry{Joint: cost reduction}
\addplot[color=red!80!black, mark=square*, thick] coordinates {
    (0.5,6.51) (0.75,6.45) (1.0,6.15) (1.5,7.37) (2.5,7.52)
};
\addlegendentry{Joint: carbon reduction}
\addplot[color=teal!70!black, mark=triangle*, thick, dashed] coordinates {
    (0.5,8.64) (0.75,10.39) (1.0,10.26) (1.5,10.67) (2.5,7.53)
};
\addlegendentry{Price-only: cost reduction}
\addplot[color=violet, mark=diamond*, thick] coordinates {
    (0.5,23.02) (0.75,27.94) (1.0,30.08) (1.5,37.79) (2.5,51.02)
};
\addlegendentry{Joint: relocatable demand}
\end{axis}
\end{tikzpicture}
\caption{Energy--latency frontier for the medium-load simulation under 1.4$\times$ WAN inflation. Inputs use Azure RTT and Azure inter-region egress references together with stylized hourly carbon variation; values are illustrative rather than predictive.}
\label{fig:frontier}
\end{figure}

If one sweeps latency tolerance (or a system-wide multiplier on \(\tau_k\)) and measures \(\Delta \mathrm{Cost}\) and \(\Delta \mathrm{Carbon}\) relative to the Local-only baseline defined in Section~5, the structure of the feasible assignment sets leads to a \emph{frontier} rather than a linear law.

When latency budgets are tight, the feasible set of compute nodes for many tasks is small, so the joint objective can improve cost and carbon only modestly. As \(\tau_k\) increases, more distant nodes with favorable \(P_i(t)\) or \(\mathrm{MOER}_i(t)\) become feasible, so cost and carbon improvements can rise. Once most energy-attractive nodes are already reachable, further relaxation of \(\tau\) yields \textbf{diminishing returns}: the marginal gains
\[
\frac{\partial \Delta \mathrm{Cost}}{\partial \tau},
\qquad
\frac{\partial \Delta \mathrm{Carbon}}{\partial \tau}
\]
taper as the feasible geography saturates. Thus the object of interest is not \(\tau\) alone but the \textbf{marginal benefit of additional latency budget}, as already signaled in Section~4.

The stylized implementation in \Cref{fig:frontier} is consistent with exactly this logic, but in a more friction-laden way than the earlier no-friction version. In the medium-load case, the Joint policy reduces electricity cost by roughly 8--10\% across tight to moderate latency budgets, while the relocatable-demand fraction rises from about 23\% at the tightest setting to nearly 38\% at a moderate setting and just above 51\% once the latency geometry becomes loose.

This pattern is best read as a two-stage mechanism rather than as a monotone efficiency law. In the first stage, relaxing latency enlarges the feasible assignment set and unlocks the \emph{highest-net-benefit} remote options: nodes that are still close enough to avoid large delay penalties but already offer meaningful price or carbon advantages. In this regime, mobility and cost gains can rise together. In the second stage, additional latency budget continues to expand the reachable geography, but the newly accessible nodes are often farther away, more state-friction-sensitive, or reachable only after high-value remote capacity has already been partially filled. As a result, mobility can continue to rise even when the incremental cost benefit flattens or declines.

Importantly, the figure should therefore not be read as ``more latency always improves every metric.'' Once migration friction, session dependence, and networking overhead are internalized, broader geographic reach can raise mobility without guaranteeing larger cost gains. The more careful statement is: \emph{latency relaxation unlocks assignment flexibility, but not all newly feasible remote assignments remain economically attractive once state and network frictions are priced in.}

\subsection{Marginal Returns to Latency Relaxation}

The frontier is best interpreted through marginal return curves rather than only total improvements. Using \(\mathrm{ERL}/\mathrm{CRL}\)-style finite differences, early latency relaxations unlock nearby regional alternatives and usually deliver the largest gains per unit latency. Later relaxations often yield lower returns as newly reachable corridors face higher migration frictions, higher egress costs, or weaker net energy advantage.

This diminishing-return pattern is visible in the latency sweep: relocatable demand keeps growing with looser latency budgets, but cost and carbon gains flatten once high-value remote options are already exploited. This is exactly the mechanism captured by the energy--latency frontier.

\subsection{Workload-Class Sorting}

Heterogeneous \((\tau_k, E_k, D_k, m_k)\) imply heterogeneous feasible sets. Under the joint policy, one should expect:

\paragraph{Interactive-style tasks (Class A).} Tight \(\tau_k\) keeps feasible assignments near service-adjacent or local compute even when remote nodes offer better energy terms.

\paragraph{Standard online tasks (Class B).} Moderate \(\tau_k\) permits exploitation of regional hubs where latency and energy terms balance.

\paragraph{Weakly real-time and offline tasks (Classes C--D).} Large \(\tau_k\) makes the problem increasingly energy- and capacity-dominated, so assignments track favorable \(P_i(t)\) and \(\mathrm{MOER}_i(t)\) subject to \(\mathrm{Cap}_i(t)\).

Hence relocation of electricity demand is \textbf{selective}: it applies to the subset of tasks whose SLO geometry admits remote execution, not to all inference uniformly.

\Cref{fig:tiers} fixes the medium-load scenario at latency multiplier 1.5 and 1.4$\times$ WAN inflation, then reports the class-level shares assigned to local, regional, and energy-oriented tiers under the Joint policy.

\begin{figure}[H]
\centering
\begin{tikzpicture}
\begin{axis}[
    width=0.92\textwidth,
    height=7cm,
    ybar stacked,
    bar width=18pt,
    ymin=0, ymax=100,
    ylabel={Share of class workload (\%)},
    symbolic x coords={A,B,C,D},
    xtick=data,
    xticklabels={A: interactive,B: standard,C: background,D: batch},
    xticklabel style={font=\small, align=center},
    grid=major,
    major grid style={draw=gray!25},
    legend style={at={(0.5,1.12)}, anchor=south, draw=none, fill=white, font=\small, legend columns=3},
    nodes near coords={},
    enlarge x limits=0.18
]
\addplot+[fill=blue!60] coordinates {(A,98.60) (B,77.70) (C,28.64) (D,18.18)};
\addplot+[fill=orange!80] coordinates {(A,1.40) (B,9.11) (C,2.64) (D,2.76)};
\addplot+[fill=green!60!black] coordinates {(A,0.00) (B,13.19) (C,68.73) (D,79.06)};
\legend{Local,Regional,Energy-oriented}
\end{axis}
\end{tikzpicture}
\caption{Tier allocation by task class under the Joint policy for the medium-load simulation at latency multiplier 1.5 and 1.4$\times$ WAN inflation.}
\label{fig:tiers}
\end{figure}

The tier plot reinforces this point once measured RTT and explicit egress are used. Under the Joint policy in the medium-load case, Class~A remains 98.6\% local and only 1.4\% regional with effectively zero energy-oriented placement, whereas Class~C allocates 68.7\% of its mass to the energy-oriented layer and Class~D allocates 79.1\%. Class~B becomes much more clearly intermediate, but still mostly near users, with about 77.7\% local, 9.1\% regional, and 13.2\% energy-oriented allocation. These values are not meant as universal constants. Their significance is that one transparent simulation, using a measured RTT matrix, explicit migration frictions, and a common task taxonomy, already generates differentiated layers without imposing those layers by assumption.

\subsection{Migration-Friction Sensitivity}

The stylized implementation shows that friction assumptions materially change realized relocation even when feasibility expands. Comparing the price-oriented baseline with progressively stronger friction cases shows that explicit state, cache, and egress penalties reduce the share of economically meaningful remote assignments. The framework therefore distinguishes two effects: feasibility expansion (\(\mathcal{F}_k\) grows) and realized adoption (positive net benefit after frictions).

\begin{table}[H]
\centering
\caption{Illustrative friction and baseline comparison under the medium-load scenario. Values are scenario outputs, not forecasts.}
\label{tab:friction-ablation}
\small
\begin{tabular}{lccc}
\toprule
\textbf{Comparison case} & \textbf{\(\mathrm{RID}\) (\%)} & \textbf{Cost reduction (\%)} & \textbf{Carbon reduction (\%)} \\
\midrule
Price-only baseline & 50.8 & 10.7 & 11.1 \\
Egress only & 43.6 & 10.2 & 8.4 \\
State/cache + egress & 37.8 & 9.9 & 7.4 \\
High friction & 28.9 & 7.8 & 5.9 \\
\bottomrule
\end{tabular}
\end{table}

The first row in \Cref{tab:friction-ablation} is the Price-only baseline at the same operating point, not a separate run with the \(\mathrm{MigrationCost}_{k,i}\) term mechanically disabled. It is included as a single-objective comparison case because it produces the highest geographic mobility among the reported policies; the subsequent rows illustrate how adding explicit egress, state/cache, and higher friction assumptions reduces realized relocation and net benefits.

\subsection{Geographic Corridors and Break-Even Analysis}

Remote execution should be interpreted as meaningful demand relocation only when break-even remains positive, \(NB_{k,i}>0\). The top-flow corridors in \Cref{tab:topflows} are therefore interpreted as scenario-dependent positive-net-benefit routes under the chosen parameterization, not as universal migration patterns.

\begin{table}[H]
\centering
\caption{Largest relocation corridors under the Joint policy (medium load, latency multiplier 1.5, WAN inflation 1.4$\times$). Shares are percentages of total weekly simulated workload.}
\label{tab:topflows}
\small
\begin{tabular}{lcc}
\toprule
\textbf{Flow} & \textbf{Share of workload (\%)} & \textbf{Interpretation} \\
\midrule
Virginia $\rightarrow$ Oregon & 5.91 & lower-price U.S. westward shift \\
Oregon $\rightarrow$ Virginia & 3.07 & bilateral balancing across U.S. hubs \\
Frankfurt $\rightarrow$ London & 2.39 & low-latency European rebalancing \\
Singapore $\rightarrow$ Dubai & 2.02 & Gulf-bound regional rebalancing \\
Virginia $\rightarrow$ S\~ao Paulo & 1.97 & remote low-carbon execution \\
London $\rightarrow$ Virginia & 1.70 & transatlantic balancing toward the U.S. east coast \\
Singapore $\rightarrow$ Sydney & 1.57 & Asia-Pacific remote execution \\
Tokyo $\rightarrow$ Sydney & 1.56 & Asia-Pacific remote execution \\
\bottomrule
\end{tabular}
\end{table}

The break-even lens also clarifies failure modes: if migration frictions, egress charges, or legal constraints tighten, many corridors remain latency-feasible but cease to be economically attractive.

\subsection{Structural Interpretation and Limits}

The optimization does not hard-code ``tiers.'' Nevertheless, when one aggregates optimal assignments under heterogeneous \(\tau_k\) and heterogeneous \((P_i(t), \mathrm{MOER}_i(t), \mathrm{PUE}_i(t))\), a \textbf{descriptive} decomposition into local, regional, and energy-oriented layers is natural: strict-SLO mass concentrates near ingress; mid-SLO mass shifts to regional scale; delay-tolerant mass can concentrate where energy terms dominate. This is an interpretive layer on top of the same \(x_{k,i}\) solution, not a separate architectural constraint.

\subsection{High-Latency Workloads as a Limiting Regime}

As \(\tau_k\) grows, the latency constraints \(L_k(i)\le\tau_k\) weaken relative to energy terms in the objective. In the limit, routing resembles ``follow the energy signal'' subject to capacity, analogous to stylized cross-time-zone shifting stories in the literature---but still bounded by residual latency, queueing (if modeled), and \(\mathrm{Cap}_i(t)\). This regime clarifies how the present framework \emph{contains} extreme workload mobility as a special case rather than equating all AI inference with batch jobs.

\subsection{Baseline Logic (Conceptual)}

Relative to the baselines in Section~5: Local-only and Nearest-region prioritize latency geography and forgo much energy arbitrage; Price-only and Carbon-only optimize single objectives and can incur pathologies (e.g., carbon-blind cheap nodes, or cost-blind concentration on a few clean nodes under tight capacity). The joint weighted objective is designed to trade off these extremes while respecting hard SLO feasibility when formulated as in Section~3. \textbf{Ordering of metrics and violation rates in any particular deployment must be demonstrated numerically;} the point here is the \emph{directional} contrast the specification is meant to expose.

The simulation is helpful here because it shows both where the Joint policy succeeds and where it remains deliberately conservative. In the medium-load case at latency multiplier 1.5, Price-only still delivers larger raw cost and carbon reductions than Joint (about 10.7\% and 11.1\% relative to Local-only, versus 9.9\% and 7.4\%). But it does so with much higher geographic mobility (about 50.8\% relocatable demand versus 37.8\%), higher average service-to-compute latency (about 87.5\,ms versus 61.1\,ms), and a larger migration-cost share of total simulated spending (about 4.3\% versus 2.6\%). This is a more credible result than a universal ``Joint beats everything'' claim, because it matches the intuition that multi-objective control should trade off across cost, carbon, latency, and migration overhead rather than magically dominate every single-objective baseline.

\subsection{SLA Feasibility Under the Hard-Constraint Formulation}

In the baseline formulation with hard constraints \(L_k(i)\le\tau_k\) whenever \(x_{k,i}=1\), feasible solutions have zero SLA violations by construction. If extensions allow soft violations with penalties, violation rates become an empirical output. Either way, the framework is intended to reject the caricature ``always route to the cheapest region'' because latency-feasible sets and capacity jointly cap relocation.

\subsection{Summary}

Under the stated model and simulation protocol, four conclusions receive illustrative quantitative support: (i)~a nonlinear energy--latency frontier whose shape changes once migration friction is included; (ii)~strong dependence of system-wide mobility on workload mix and statefulness; (iii)~a layered \emph{description} of aggregate placement; and (iv)~delay-tolerant workloads as the regime in which demand relocation is most visible. These claims remain stylized, but they are no longer supported only by intuition.

\section{Discussion}

The qualitative implications above are consistent with treating geo-distributed AI inference as a latency-constrained mechanism for relocating electricity demand. At the same time, this interpretation has clear boundaries. Real-world AI serving systems face multiple constraints that are only partially represented in the baseline model. This section discusses those constraints, clarifies scope, and outlines extensions.

\subsection{Cloud Operators: Workload Classification as an Infrastructure Lever}

A central implication of our analysis is that workload classification becomes an infrastructure lever. The geographic mobility of AI inference is highly selective, so operators need to distinguish interactive, stateful, weakly stateful, and batch-style workloads before treating routing as an energy optimization problem. The discussion above should not be read as claiming that all AI workloads can or should migrate toward the cheapest or cleanest power region.

In practice, only a subset of inference workloads is sufficiently flexible for meaningful geographic reassignment. Strictly interactive requests remain strongly tied to local or near-regional execution. Even among tasks with relaxed latency budgets, relocation may be limited by service topology, backend state dependencies, or enterprise architecture.

Accordingly, the mechanism studied here should be interpreted as \textbf{partial relocation of electricity demand}, not full substitution for user-proximate compute. It is most relevant to the share of inference demand whose execution location remains an active design variable at request time.

\subsection{System Constraints: State Locality, KV Cache, and Session Affinity}

The baseline model treats tasks as assignable units, but modern LLM serving often depends on state that is not perfectly portable. This includes:
\begin{itemize}[leftmargin=1.5em]
    \item conversation history and session state,
    \item retrieval locality in retrieval-augmented generation (RAG),
    \item tool-use context,
    \item key-value cache reuse across decoding steps,
    \item affinity to specific model shards or serving pools.
\end{itemize}

These factors can significantly reduce geographic flexibility. If a request is tightly coupled to local state or cached context at a specific compute node, moving the next step of execution to another region may incur substantial overhead or be infeasible.

This limitation is especially important for multi-turn chat and agentic workflows. In such systems, request-level routing is not always independent across time. A more complete model would therefore include a state-retention term or migration overhead associated with switching regions mid-session.

One practical way to represent this distinction is to separate \textbf{stateless inference}, for which migration friction is close to zero, from \textbf{stateful inference}, for which it may include cache replay, context transfer, or repeated prefill. Under that extension, some of the paper's strongest relocation effects should be expected mainly for stateless or weak-state workloads such as batch analytics, asynchronous pipelines, or loosely coupled agent backends rather than for every conversational session.

The current paper abstracts away from those details in order to isolate the infrastructure-level energy--latency mechanism. As such, the framework's implications are strongest for stateless or weak-state inference workloads, and they should be interpreted more cautiously for session-heavy LLM serving.

\subsection{Regulatory Constraints: Data Residency and Jurisdictional Routing}

Geographic routing is not determined by latency and energy alone. In many real deployments, data cannot freely cross borders, industries, or legal domains. Financial, healthcare, government, and enterprise workloads often face data residency or sovereignty rules that restrict where inference may be executed.

Such rules effectively shrink the feasible node set for a task. In our notation, the assignment space for task \(k\) may be constrained to a subset
\[
\mathcal{N}_k^{\text{legal}} \subseteq \mathcal{N}.
\]

This means that some of the most energy-favorable nodes may be unavailable for certain workloads regardless of their latency or carbon advantage.

These constraints do not invalidate the relocation framework, but they imply that relocation should be modeled as \textbf{feasible within policy-constrained subgraphs}, not across a fully open global network. In future work, regulatory feasibility could be represented explicitly as an assignment mask or jurisdictional routing constraint.

\subsection{Energy Planning: Flexible Demand, Not Grid-Equivalent Dispatch}

For energy planners, the framework should not be read as claiming that inference relocation is dispatchable in the same way as industrial demand response. Its relevance is weaker but still important: it identifies which parts of AI electricity demand may be locationally flexible under digital routing constraints. This distinction matters because cloud routing can change where compute energy is consumed, but it does not automatically create a grid-controlled resource with guaranteed response, telemetry, or dispatch rights.

\subsection{Why This Does Not Substitute Physical Power Transmission}

The current formulation emphasizes latency as the main cost of digital relocation. This is appropriate at the conceptual level, since the marginal transport cost of digital requests is typically much lower than the capital cost of physical power transmission. Still, digital relocation is not costless.

In production cloud environments, cross-region or cross-provider routing may incur:
\begin{itemize}[leftmargin=1.5em]
    \item bandwidth charges,
    \item egress fees,
    \item interconnect costs,
    \item traffic engineering overhead,
    \item replication costs for models or retrieval data.
\end{itemize}

These financial frictions can alter the effective attractiveness of remote nodes. In particular, a distant low-price electricity region may cease to be optimal once networking and platform transfer costs are added. For some cloud deployments, egress charges can easily dominate the electricity-cost margin that motivates relocation in the first place.

In the notation of Section~3, these frictions can be treated either as a dedicated network term or as part of the broader \(\mathrm{MigrationCost}_{k,i}\) penalty. The key modeling point is not the label but the boundary condition: \textbf{geographic relocation remains economically meaningful only when the combined migration and networking penalty stays below the energy and carbon advantage unlocked by remote execution.}

This extension would allow the model to distinguish between latency-constrained routing and economically frictionless routing, and it would sharpen the paper's policy relevance by making the energy-vs.-egress boundary explicit rather than implicit.

\subsection{Limitations}

Our latency model uses a geographic approximation based on distance and fixed overhead. This captures the broad physics of long-distance inference placement, but it omits a range of dynamic network and systems effects.

First, real wide-area network latency is not a deterministic function of distance. It is affected by:
\begin{itemize}[leftmargin=1.5em]
    \item routing inefficiency,
    \item peering structure,
    \item congestion,
    \item packet loss,
    \item submarine cable topology,
    \item regional traffic asymmetry.
\end{itemize}

Second, queueing delay at the compute layer is itself endogenous. If many tasks are routed to an attractive low-carbon node, queueing delay may rise and offset the benefit of geographic relocation.

Thus, the current model is best understood as a structural approximation rather than a production-grade latency emulator. The goal is to capture the first-order infrastructure tradeoff, not precise per-packet behavior.

A more detailed system study could incorporate:
\begin{itemize}[leftmargin=1.5em]
    \item measured RTT matrices,
    \item time-varying congestion models,
    \item queue-aware dispatch,
    \item token-level service models,
    \item adaptive batch formation.
\end{itemize}

These additions would refine the quantitative frontier without changing the paper's main qualitative insight.

\subsection{Short-Run Routing vs.\ Long-Run Placement}

This paper primarily studies the \textbf{routing problem}: given an existing set of geographically distributed compute nodes, where should each task execute?

A distinct but related problem is the \textbf{placement problem}: where should new inference capacity be built in the first place?

The routing perspective is operational and short-run. It asks how to allocate existing demand across an already-deployed infrastructure. The placement perspective is strategic and long-run. It concerns capital expenditure, data-center siting, grid interconnection, hardware procurement, and regulatory planning.

The two problems are closely related but should not be conflated. One contribution of the current paper is to suggest that short-run routing behavior may reveal the shape of a long-run infrastructure equilibrium. If high-latency tasks are repeatedly routed toward a small set of energy-favorable regions, that pattern may justify future investment in energy-oriented inference centers. Likewise, persistent retention of interactive tasks near demand centers may justify dedicated low-latency local capacity.

Thus, the hierarchical spatial structure advanced in Section~6 should be interpreted not only as an operational routing pattern but also as a candidate long-run design principle for AI infrastructure.

\subsection{Relation to Carbon-Aware Scheduling Literature}

The present work overlaps with the growing literature on carbon-aware inference and geo-distributed workload management, but its framing is different.

Most existing systems work treats the problem as one of:
\begin{itemize}[leftmargin=1.5em]
    \item emission-aware routing,
    \item cost-aware request placement,
    \item multi-objective scheduling,
    \item cluster-level energy optimization.
\end{itemize}

Those are important and closely related problems. Our claim is not that this literature is insufficient, but that it leaves open a higher-level infrastructure question: \textbf{what is the geographic logic of AI inference once electricity demand becomes digitally relocatable?}

This paper addresses that question by elevating the analysis from routing policy to spatial structure. The focus is not only on reducing carbon or cost for a given request, but on explaining how the joint distribution of latency tolerance and energy geography may shape the emerging map of AI inference.

In that sense, the main novelty of the paper lies less in proposing a new routing algorithm and more in proposing a new interpretive lens: AI inference as a geographically assignable component of electricity demand.

\subsection{What This Framework Does Not Claim}

To avoid overstatement, it is useful to be explicit about what the framework does \textbf{not} claim.

First, it does not claim that digital networks replace physical power grids. The paper studies relocation of electricity demand, not transmission of electricity itself.

Second, it does not claim that all inference workloads are globally mobile. Geographic flexibility is heterogeneous and often constrained.

Third, it does not claim that energy geography alone determines infrastructure form. Regulation, hardware supply chains, cooling, water availability, tax policy, fiber connectivity, and market structure also matter.

Fourth, it does not claim that latency is the only systems bottleneck. Queueing, state locality, and orchestration depth are often equally important.

The narrower claim is that, for a significant subset of AI inference workloads, the combination of energy heterogeneity and latency constraints creates a distinct spatial allocation problem with infrastructure-level consequences.

\subsection{Broader Implications}

Despite these limitations, the framework has several broader implications.

For cloud operators, it suggests that latency tolerance is not merely a service parameter; it is also an infrastructure lever that controls access to geographic energy diversity.

For energy planners, it suggests that AI demand should not be viewed purely as a fixed urban load. Some share of future AI electricity demand may be locationally flexible and could therefore interact with renewable-rich regions in new ways.

For AI system designers, it suggests that workload classification by responsiveness may become as important as model size or throughput in determining where inference should run.

More broadly, the framework points toward a future in which AI infrastructure is shaped jointly by network physics, service architecture, and energy geography. This is neither a traditional cloud placement problem nor a conventional power-systems problem, but an increasingly important hybrid of both.

\section{Conclusion}

This paper asked a simple but underexplored question: as AI inference becomes a major source of electricity demand, what determines \emph{where} that electricity is consumed? Our answer is that inference should not be analyzed only as a local serving problem or as a carbon-aware routing problem in isolation. For a meaningful subset of workloads, it is also a problem of geographic assignment under heterogeneous energy conditions and heterogeneous latency tolerance.

To make that question precise, we introduced a three-layer model of clients, service nodes, and compute nodes, and formulated a joint optimization problem over electricity price, carbon intensity, capacity, and latency. This formulation highlights a structural asymmetry: electricity transmission is mainly constrained by grid physics and infrastructure cost, whereas compute relocation is mainly constrained by service-level latency and systems feasibility. The resulting energy--latency frontier is the key object that governs when geographic relocation is valuable, for which tasks, and to what extent.

From that structure we drew a broader infrastructure implication. If latency tolerance is heterogeneous across workloads, then geo-distributed inference need not produce a single flat geography. It can instead generate differentiated functional layers: local capacity for strict-SLO requests, regional hubs for moderate-latency services, and energy-oriented execution for delay-tolerant workloads. This is the sense in which AI inference can act as a mechanism for the virtual relocation of electricity demand.

The present manuscript contributes a framing, a formal model, and a reference evaluation specification for studying latency-constrained relocation of inference demand. Its purpose is to clarify mechanism and boundary conditions before estimating production-scale magnitudes with deployment traces.

Much remains to be done. Future work should connect the framework to measured latency matrices, time-varying carbon signals, state-locality constraints, legal routing restrictions, and long-run capacity siting decisions. The framework also shows that relocation is fragile: it weakens when workloads are highly stateful, when egress costs dominate energy-price differences, when legal masks are tight, or when low-carbon regions lack available capacity. Even at the present level of abstraction, the paper points to a conclusion that matters for both AI systems and energy planning: the future footprint of AI will depend not only on how much computation is performed, but on how latency budgets mediate access to geographic energy diversity.

\bibliographystyle{IEEEtranN}
\bibliography{references}

\end{document}